\documentclass[11pt]{article}
\usepackage{arxiv}
\usepackage{booktabs}
\usepackage{amsmath}
\usepackage{graphicx}
\usepackage{url}
\usepackage[numbers]{natbib}

\title{Reliable Neural-Codec Text-to-Speech by ASR Self-Verification and Distillation:\\ Near-Zero Catastrophic Failures Across Models and Codecs}
\author{%
  Ali Asaria \\ Transformer Lab \and
  Tony Salomone \\ Transformer Lab \and
  Deep Gandhi\thanks{Corresponding author: \texttt{deep@lab.cloud}} \\ Transformer Lab
}
\date{}
\runningtitle{Reliable codec-TTS by self-verification}

\begin{document}
\maketitle

\begin{abstract}
Open autoregressive neural-codec text-to-speech (TTS) models sound excellent on
typical inputs yet suffer \emph{stochastic catastrophic failures}: on a meaningful
fraction of utterances they emit silence, terminate early, or collapse into repetitive
or hallucinated content, a deployment-disqualifying behaviour. We show this failure
mode is cheap to remove. Under a single format-robust metric (a catastrophic-failure
rate via an ASR round-trip), \textbf{best-of-$N$ ASR self-verification drives failures
to near-zero}: no observed failures remain by $N{=}2$ on a standard corpus
(LibriSpeech, 95\% upper bound $0.008$) and by $N{=}4$ on a hard prompt set. Crucially,
this is not an artifact of one model, the reduction \textbf{replicates across four open
codec-TTS systems and three different neural codecs} (XCodec2, SNAC, Mimi), reaching the
near-zero floor by $N{=}2$ on three of the four (a larger Llasa, whose numbers carry a
generation-budget caveat, being the exception). We
then make the fix free at inference time: \textbf{distilling} the self-verified
behaviour into the model recovers much of the robustness in single-shot decoding, closing
$\sim$52--58\% of the failure mass on hard inputs ($0.199\rightarrow0.083$--$0.096$) at no
test-time cost (best-of-$N$ itself is a reference-aware oracle, so the distilled
single-shot rate is the deployable number). The distillation gain concentrates where it
is needed (hard inputs); on already-reliable prose there is no headroom and no
detectable change, and where the base is already near the floor distillation can even
regress slightly. A controlled comparison adds a clean negative:
offline direct preference optimization (DPO/IPO) does not beat plain supervised
distillation, and an online iterative variant is promising but not statistically
separable at our evaluation size. Throughout we give an honest account of the one model
that resists (a larger Llasa where scale did not obviously help) and of a rare-word
capability ceiling that no self-distillation method overcomes.
\end{abstract}

\section{Introduction}
\label{sec:intro}
Neural-codec TTS, an autoregressive language model over discrete acoustic tokens,
decoded by a neural codec, has become a dominant open paradigm. Its weakness is
reliability: the same model that renders most sentences cleanly will, unpredictably,
fail catastrophically on others. On a broad evaluation of Llasa-1B that includes
deliberately adversarial rare-word and number/date prompts, we measure a
catastrophic-failure rate of $\sim$0.36; on a hand-written hard set used for our
robustness experiments (excluding the rare-word capability ceiling, \S\ref{sec:setup})
the single-shot rate is $\sim$0.27. Either way, a large fraction of generations is
a dropout (silence or early stop) or a collapse (repetition or wrong content). For
deployment this is disqualifying, and it is precisely the ``skips, repeats, and
attention-collapse hallucinations'' that motivate robustness post-training.

Our message is that this problem is \emph{cheap to fix}. A simple test-time procedure
that samples a few candidates and keeps an ASR-verified one removes catastrophic
failures whenever the model is capable of a good sample at all, and it does so not on
one model but across architectures and codecs, which is what makes it credible as a
general fix rather than an idiosyncrasy. The only cost is $N\times$ inference, and we
remove even that by \emph{distilling} the verified behaviour back into the model so
that ordinary single-shot decoding inherits it. That transfer is most effective exactly
where it matters: the hard inputs that account for the failure budget. Where the base
model is already reliable, there is simply nothing to recover.

Our contributions:
\begin{enumerate}
\item \textbf{A general fix at test time.} Best-of-$N$ ASR self-verification reduces
catastrophic failures across \textbf{four models and three neural codecs} (XCodec2,
SNAC, Mimi), reaching the rule-of-three zero floor on three of four
($0.27\rightarrow$ none by $N{=}4$ on hard inputs; $0.06\rightarrow$ none by $N{=}2$ on
standard prose) and reducing (though not eliminating) failures on the fourth, whose
rates carry a generation-budget caveat, the effect is not model-specific
(\S\ref{sec:results-bon}). %
\item \textbf{Making the fix free at inference.} A single distillation pass on
self-verified samples recovers much of the robustness in single-shot decoding on hard
inputs ($0.199\rightarrow0.083$--$0.096$, $\sim$52--58\% of the failure mass) at no
test-time cost; the gain concentrates on hard inputs and is null where the base is
already reliable (\S\ref{sec:results-distill}). %
\item \textbf{A controlled comparison} that yields a clean negative: offline DPO/IPO do
not beat supervised distillation, and an online iterative variant is the most promising
direction but not statistically separable at our scale (\S\ref{sec:results-pref}). %
\item \textbf{A format-robust metric} for codec-TTS, and the finding that number/date
robustness cannot be measured by WER (a digit-vs-word confound), motivating a
dropout/collapse rate (\S\ref{sec:setup}, \S\ref{sec:analysis}). %
\end{enumerate}

\section{Related Work}
\label{sec:related}
Preference optimization of TTS via automatic rewards (ASR-WER, speaker similarity, MOS
predictors) has been applied to AR codec models in CosyVoice~2~\citep{2412.10117v3},
Koel-TTS~\citep{2502.05236v2}, and TTS-1~\citep{2507.21138v1}; online RL (GRPO) variants
appear in Align2Speak~\citep{2509.21718v1} and multi-metric
alignment~\citep{2508.17229v2}. Surveys of preference learning for
audio~\citep{2511.13936v1} and of direct alignment
algorithms~\citep{2502.01237v3,2503.11701v1,2410.15595v3} repeatedly note that
\emph{offline} DPO underperforms supervised fine-tuning or online methods, a pattern
our controlled comparison reproduces and that motivates our reading of the online,
iterative variant as the promising direction. Robustness failure modes of AR codec-LMs
are characterised by \citet{2502.18924v4,2502.03128v1}, and evaluation tooling (MOS
prediction, TTSDS) is developed by \citet{2409.09305v1,2407.12707v3}. Unlike prior work, which
typically applies a single preference method to its own model, we (i) establish that a
\emph{test-time} verification baseline, distilled, is a strong and simple lever, (ii)
show it generalizes across four models and three codecs, and (iii) characterise
\emph{when} the distilled gain materialises.

\section{Setup}
\label{sec:setup}
\textbf{Models and codecs.} Our primary model is Llasa-1B~\citep{llasa2025}, a
LLaMA-style LM over XCodec2 speech tokens, adapted with LoRA on the released checkpoint.
For the generalization study (\S\ref{sec:results-bon}) we add three further open systems
spanning three codecs: Orpheus-3B~\citep{orpheus2025} (LLaMA over the SNAC codec),
CSM-1B~\citep{csm2025} (Sesame, over the Mimi RVQ codec), and Llasa-3B~\citep{llasa2025}
(XCodec2, a scale point). (We initially targeted XTTS-v2 but its
fine-tuning tooling proved unworkable in our environment.)

\textbf{Metric.} We score each generation by a Whisper ASR round-trip. A
\emph{catastrophic failure} is a dropout (speech-token count $<25$, or ASR returns
$\le 1$ word) or a content failure (WER $>0.5$ against the reference text). Formally,
for a generation $g$ of reference text $x$, let $n(g)$ be its decoded speech-token
count, $h(g)$ its ASR transcript, $|h(g)|$ the transcript word count, and
$\mathrm{WER}(h,x)$ the word error rate. The per-generation catastrophic-failure
indicator is
\begin{equation}
\phi(g)=\mathbf{1}\!\Big[\,\underbrace{n(g)<\tau_{\mathrm{tok}} \;\lor\; |h(g)|\le\tau_{w}}_{\text{dropout}}\;\lor\;\underbrace{\mathrm{WER}(h(g),x)>\theta}_{\text{collapse / wrong content}}\,\Big],
\label{eq:phi}
\end{equation}
with thresholds $\tau_{\mathrm{tok}}{=}25$, $\tau_{w}{=}1$ word, $\theta{=}0.5$. For a
held-out prompt set $\mathcal{P}$ with $N$ sampled candidates $g_{p,1},\dots,g_{p,N}$
per prompt, best-of-$N$ self-verification counts a prompt as failed only when
\emph{all} $N$ candidates fail, giving the catastrophic-failure rate
\begin{equation}
\mathrm{CFR}_N=\frac{1}{|\mathcal{P}|}\sum_{p\in\mathcal{P}}\;\prod_{k=1}^{N}\phi(g_{p,k}),
\label{eq:cfr}
\end{equation}
so $\mathrm{CFR}_N$ is non-increasing in $N$ and reaches $0$ once every prompt has at
least one non-catastrophic candidate within $N$ draws; the single-shot rate is the
$N{=}1$ case, $\mathrm{CFR}_1$. We adopt
this format-robust definition after finding that number/date WER is dominated by
digit-versus-word formatting noise, $\sim$0.15 in both raw and pre-normalized
conditions, with the effect of interest buried under canonicalization artifacts
; we therefore demote numbers/dates to qualitative reporting and centre the metric
on the dropout/collapse rate.

\textbf{Evaluation sets.} We use two complementary held-out sets, both disjoint from
training by text. (i) A \emph{hard set}: hand-written prompts spanning normal, long,
and tongue-twister buckets, inputs chosen to stress robustness, reported over 26
prompts $\times$ 6 generations (156). Rare-word and number/date buckets are analysed
separately as capability and measurement limits respectively. (ii) A \emph{standard
corpus} for external validity: 120 held-out LibriSpeech test-clean transcripts $\times$
3 generations (360), with training prompts from the disjoint dev-clean split. The
generalization study uses this corpus for all four models.

To avoid confusion between several related base rates, we fix terminology: the
\emph{broad-eval} rate ($0.36$) includes the adversarial rare-word and number/date
buckets; the \emph{hard-set} single-shot rate excludes the rare-word capability ceiling
and is $0.269$ at best-of-1 and $0.199$ on the larger 156-generation distillation
eval; we anchor all before/after distillation comparisons on the latter ($0.199$).

\textbf{Statistical reporting.} Failure rates are binomial proportions over the
generation counts above. We report Wilson 95\% confidence intervals for proportions
that carry a claim; cells with zero observed failures are reported with a rule-of-three
95\% upper bound ($3/n$) rather than as a true zero. The shared-prompt SFT-versus-DPO
contrast is read as a paired comparison. All evaluations are single training runs per
condition; we therefore avoid ranking methods whose intervals overlap.

\section{Method}
\label{sec:method}
\textbf{Best-of-$N$ self-verification (test time).} Sample $N$ candidates for a prompt,
transcribe each with ASR, and return the lowest-WER non-dropout candidate. A prompt
``fails'' at best-of-$k$ only if all $k$ of its candidates are catastrophic
(Eq.~\ref{eq:cfr}). The smallest $N$ at which a prompt stops failing is a per-input
difficulty signal.

\textbf{Distilling self-verification into single-shot inference.} To avoid the
$N{\times}$ inference cost, we fine-tune the model (LoRA) on the best self-verified
sample per training prompt, so single-shot decoding imitates the verified behaviour.
(Where the base failure rate is already near the floor, Orpheus $0.008\rightarrow0.033$,
Llasa-3B $0.108\rightarrow0.142$, distillation has nothing to recover and can
regress slightly; the gain below is on the hard set, which has room.) We compare plain
supervised distillation (SFT on the chosen sample) against preference
optimization that additionally uses a rejected sample:
\begin{equation}
\mathcal{L}_{\text{DPO}} = -\log\sigma\!\big(\beta\,[(\pi_\theta{-}\pi_{\text{ref}})_{c} - (\pi_\theta{-}\pi_{\text{ref}})_{r}]\big),
\end{equation}
with $\pi_{\text{ref}}$ the LoRA-disabled base. We also test IPO and a
\emph{failure-targeted} pairing (FTPO) in which the rejected sample is specifically a
dropout/collapse. Finally, an \emph{online iterative} variant makes the pairs
on-policy: after each round we re-sample from the current policy, re-score by ASR,
rebuild chosen/rejected pairs, and continue training (three rounds).

\section{Results}
\label{sec:results}

\subsection{Self-verification drives failures to near-zero, across four models and three codecs}
\label{sec:results-bon}
\textbf{The fix works, and it works everywhere we tried it.} On our primary model,
ASR self-verification removes catastrophic failures once a handful of candidates are
drawn: on the hard set the rate falls from $0.269$ single-shot ($95\%$ CI
$[0.205,0.343]$) to $0.038$ at $N{=}3$ and to no observed failures at $N{\ge}4$
($95\%$ upper bound $3/156{\approx}0.019$); on LibriSpeech it falls from $0.058$
to no observed failures already at $N{=}2$ (upper bound $0.008$)
(Table~\ref{tab:bon}).

\begin{figure}[t]\centering
\includegraphics[width=0.66\linewidth]{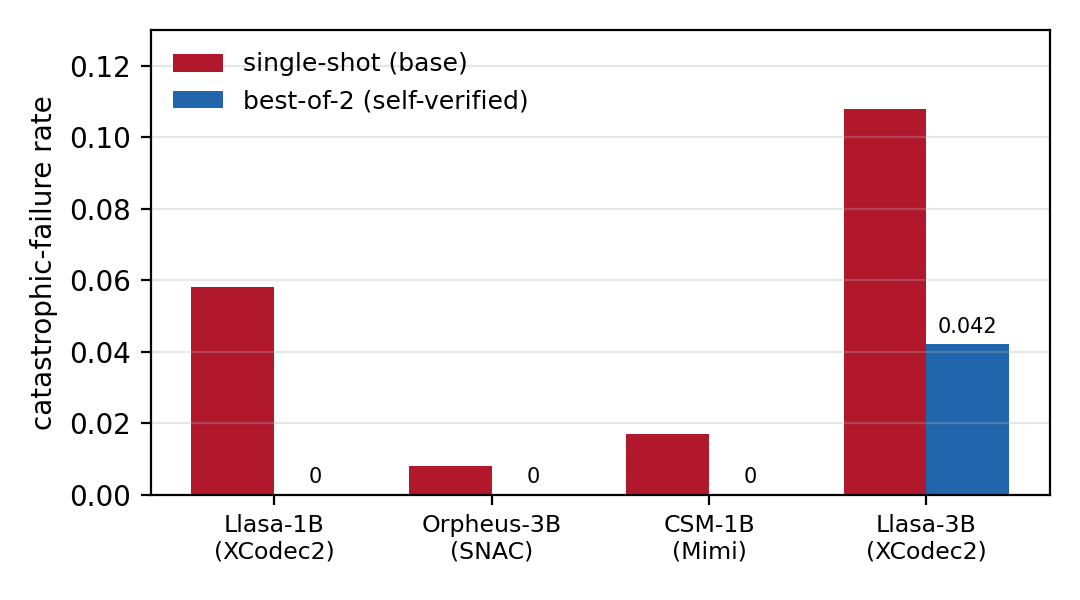}
\caption{Single-shot (base) vs.\ best-of-2 catastrophic-failure rate across four
codec-TTS models and three codecs. best-of-2 reaches no observed failures on three of
four models; Llasa-3B is the lone exception. $0.000$ cells are 95\% upper-bounded by
$0.008$.}%
\label{fig:multimodel}
\end{figure}

The result is not specific to one model. Running the identical test-time protocol
(LibriSpeech test-clean, 120$\times$3) on three further systems spanning three neural
codecs, best-of-$N$ reduces catastrophic failures on \emph{every} one
(Table~\ref{tab:multimodel}, Fig.~\ref{fig:multimodel}): single-shot rates of $0.008$
(Orpheus-3B~\citep{orpheus2025}, SNAC), $0.017$ (CSM-1B~\citep{csm2025}, Mimi)
and $0.058$ (Llasa-1B, XCodec2)
all fall to the rule-of-three zero floor by $N{=}2$, and Llasa-3B falls from $0.108$ to
$0.033$. The same simple verifier works on every model we tested, across
single-stream (XCodec2), flattened-RVQ (SNAC) and hierarchical-RVQ (Mimi) token
representations and three model families. This is evidence that best-of-$N$ reduces
catastrophic failures broadly across codec-TTS, rather than for one system, though we
note two limits on how far to read it: the rare-word capability ceiling
(\S\ref{sec:analysis}) marks failures that sampling provably cannot fix, and Llasa-3B
does not reach the floor.

\begin{table}[t]\centering
\caption{Best-of-$N$ across four open codec-TTS models and three neural codecs
(LibriSpeech test-clean, 120$\times$3). $0.000$ = no observed failures (95\% upper
bound $3/360{\approx}0.008$). ``--'' = not run.}%
\label{tab:multimodel}
\begin{tabular}{llcccc}\toprule
model & codec & base & best-of-2 & best-of-3 & distilled\\\midrule
Llasa-1B & XCodec2 & 0.058 & \textbf{0.000} & 0.000 & 0.058\\
Orpheus-3B & SNAC & 0.008 & \textbf{0.000} & 0.000 & 0.033\\
CSM-1B & Mimi & 0.017 & \textbf{0.000} & 0.000 & --\\
Llasa-3B & XCodec2 & 0.108 & 0.042 & 0.033 & 0.142\\\bottomrule
\end{tabular}
\end{table}

\textbf{Two honest qualifications.} First, the verifier is \emph{reference-aware}: it
uses the target transcript to score candidates, so these are oracle-selection upper
bounds on what self-verification can buy, the deployable, reference-free number is the
distilled single-shot column (\S\ref{sec:results-distill}), not the best-of-$N$ column.
Second, \textbf{Llasa-3B is the one model that does not reach the floor}: scaling Llasa
1B$\rightarrow$3B did not visibly improve robustness in our run (its base rate is
\emph{higher} and best-of-$N$ plateaus at $0.033$ rather than reaching zero). This carries a comparability
caveat, the 3B run used a tighter generation budget (\texttt{max\_new\_tokens}$=1024$,
added to bound a suspected hang) than the 1B run, which can truncate some longer
generations into failures and inflate its rates, so we do not read the 1B-vs-3B gap as
a clean scale effect. (We report only base and best-of-$N$ for CSM-1B: its
two-transformer architecture does not expose the single token stream our distillation
path trains on; distillation-on-CSM is future work.)

\begin{table}[t]\centering
\caption{Best-of-$N$ curve on the hard set and LibriSpeech (primary model). $0.000$ =
no observed failures (95\% upper bound $3/n$). ``, '' = not sampled.}%
\label{tab:bon}
\begin{tabular}{lcccccc}\toprule
$N$ & 1 & 2 & 3 & 4 & 5 & 6\\\midrule
hard set & 0.269 & 0.154 & 0.038 & 0.000 & 0.000 & 0.000\\
LibriSpeech & 0.058 & 0.000 & 0.000 & ,  & ,  & , \\\bottomrule
\end{tabular}
\end{table}

\subsection{Distillation makes the fix free, where it is needed}
\label{sec:results-distill}
\begin{figure}[t]\centering
\includegraphics[width=0.6\linewidth]{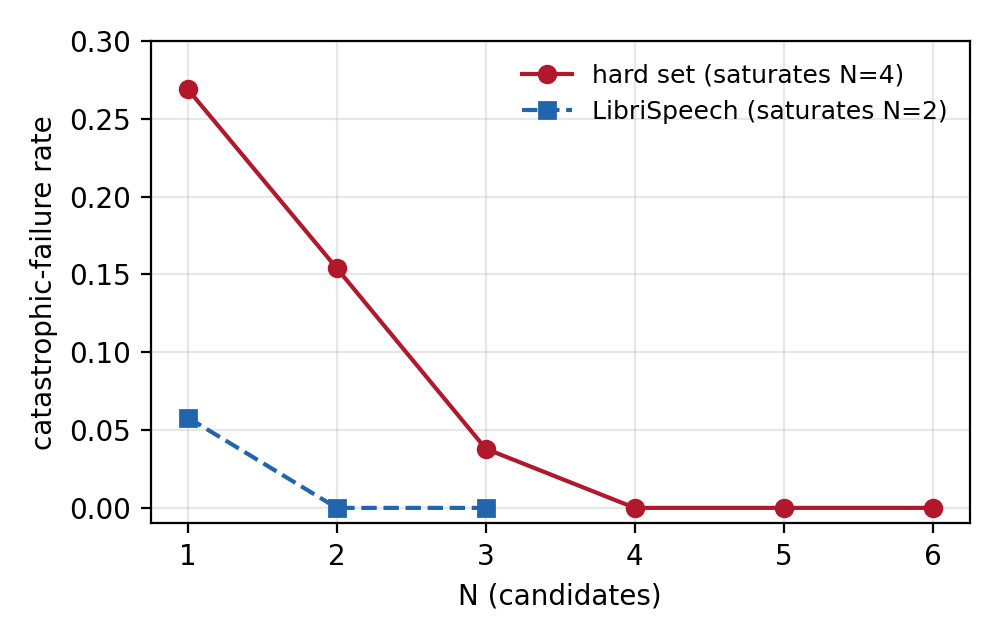}
\caption{Best-of-$N$ failure rate by input difficulty (primary model). Hard inputs
saturate at $N{=}4$, easy prose at $N{=}2$; the saturation point is the difficulty
proxy associated with distillation headroom. $x$-axis $N$, $y$-axis
catastrophic-failure rate.}%
\label{fig:bon_curve}
\end{figure}

A single distillation pass on self-verified samples transfers the robustness into
single-shot decoding at no inference cost, and the gain concentrates on the hard inputs
that dominate the failure budget (Table~\ref{tab:main}, Fig.~\ref{fig:bon_curve}). On
the hard set, distillation moves the single-shot failure rate from $0.199$
($[0.143,0.270]$) to $0.083$ (DPO, $[0.049,0.137]$) and $0.096$ (SFT, $[0.059,0.152]$)
, closing $\sim$52\% (SFT) to $\sim$58\% (DPO) of the failure mass, with a
non-trivial residual remaining. On easy LibriSpeech prose, where the base is already at $0.058$
($[0.038,0.087]$) and best-of-$N$ saturates at $N{=}2$, there is no headroom and we see
no detectable change ($0.058\rightarrow0.058$; the 95\% interval on the difference spans
zero). In short, distilled self-verification is most valuable precisely where the
model fails most. We return in \S\ref{sec:limits} to why difficulty, though the most
plausible reading, is not fully isolated from corpus source in this two-regime
comparison.

\begin{table}[t]\centering
\caption{Single-shot catastrophic-failure rate before and after one distillation pass,
by input difficulty (Wilson 95\% CIs). The measurable gain appears on the hard,
high-saturation-$N$ set and not on easy prose.}%
\label{tab:main}
\begin{tabular}{lccc}\toprule
inputs & base single-shot & saturation $N$ & distilled single-shot\\\midrule
hard set & 0.199 $[0.143,0.270]$ & $N{=}4$ & \textbf{0.083} $[0.049,0.137]$ / 0.096 $[0.059,0.152]$\\
LibriSpeech & 0.058 $[0.038,0.087]$ & $N{=}2$ & 0.058 (no detectable change)\\\bottomrule
\end{tabular}
\end{table}

\subsection{Preference optimization adds nothing over supervised distillation}
\label{sec:results-pref}
We next ask whether the preference signal (a rejected sample) helps beyond plain
supervised distillation, and find it does not. In an offline sweep on the hard set, the
preference variants (DPO $0.292$, FTPO $0.292$, IPO $0.319$) did not improve on a single
round of SFT-on-best ($0.264$); with these proportions over $\sim$108
generations their intervals overlap heavily, so we read this as ``no method beat SFT''
rather than a strict ranking. In the matched distillation comparison, SFT ($0.096$) and
DPO ($0.083$) differ by $0.013$, well inside the standard error of the difference
($\approx0.032$ on 156 shared-prompt generations), a tie.

The \textbf{online, iterative on-policy} variant gives the lowest failure rate we
observe, $0.013$ (DPO) / $0.026$ (SFT), directionally consistent with reports that
online/iterative preference optimization outperforms offline
DPO~\citep{2502.01237v3,2511.13936v1}. We do not overclaim it: those rates are $1/78$ and $2/78$ on a smaller evaluation, with
$95\%$ intervals (e.g.\ $[0.002,0.069]$) that overlap the single-pass distillation
numbers, and the value is a post-hoc minimum over many conditions. We therefore report
online iterative preference optimization as the most promising \emph{direction} for
closing the residual hard-input gap, not a confirmed improvement.

\section{Analysis}
\label{sec:analysis}
\textbf{A mechanism for the difficulty pattern.} A natural account of why the measurable
gain concentrates on hard inputs is that distillation teaches single-shot decoding to
land on the good modes the base model only reaches with several samples. When the base
is already near those modes (easy prose, small saturation $N$), there is little to move;
when it is frequently off them but a good mode exists (hard inputs, large saturation
$N$), imitating the self-verified sample collapses much of the failure mass. Under this
account the best-of-$N$ saturation point is both a difficulty measure and a proxy for
distillation headroom. We stress this is an interpretation: with two regimes it is
suggested by, not proven from, our data (\S\ref{sec:limits}).

\textbf{Failure-mode taxonomy.} Four patterns recur: (i) silence/dropout
($\sim$4.6\% of base generations); (ii) repetitive collapse (e.g.\ a looped
syllable); (iii) content hallucination (e.g.\ a short unrelated phrase for a
tongue-twister); (iv) rare-word mispronunciation.

\textbf{Capability ceiling.} Rare words (e.g.\ \emph{otorhinolaryngologist}) are a
genuine limit rather than a sampling problem: only 2 of 10 such prompts ever yield a
faithful sample within $N{=}3$. Because best-of-$N$ and distillation can only
recover prompts for which a good sample exists, both plateau high on this bucket. We
report this as a limitation, not a fixable robustness issue, no amount of
self-distillation manufactures a capability the base lacks.

\textbf{Measurement of numbers/dates.} WER cannot measure number/date robustness:
digit-versus-word canonicalization noise of $\sim$0.14 swamps the effect, and a
raw-versus-normalized A/B leaves the WER essentially unchanged (0.155 vs.\ 0.143),
i.e.\ artifact-dominated. The symptom is consistent with a failure at
\emph{tokenization} rather than acoustics; we re-centred the metric on the
dropout/collapse rate accordingly.

\section{Discussion and Limitations}
\label{sec:limits}
The practical message, catastrophic failures are cheap to remove at test time and, on
hard inputs, free at inference after a single distillation pass, comes with clear
boundaries we state plainly.

\textbf{Threats to validity.} \emph{Internal (the central caveat):} the
difficulty-dependence of the distillation gain rests on an across-regime comparison
(hard set vs.\ LibriSpeech), and those regimes differ in corpus source, hand-authored
vs.\ natural prose, length and word-frequency distribution, speaker reference, as well
as difficulty. The clean control, which we did not run, is a \emph{within-corpus,
per-prompt difficulty sweep} (stratify a single corpus by each prompt's best-of-$N$
saturation $N^\ast$ and test whether the distillation gain rises with $N^\ast$ holding
source fixed); that experiment would convert the present two-point contrast into a
dose-response relationship and is the most important next step. \emph{External:} the
test-time best-of-$N$ result spans four models and three codecs
(\S\ref{sec:results-bon}), but the \emph{distillation} analysis is shown on fewer
(Llasa-1B/3B and Orpheus); evaluation is a single language and two corpora, and the null
on easy prose may not transfer to noisier real-world text; the Llasa-3B point carries a
generation-budget caveat. \emph{Construct:} the metric is an ASR round-trip, inheriting
the ASR model's errors (we did not measure its false-positive floor on ground-truth
audio); best-of-$N$ is reference-aware and so upper-bounds rather than measures
deployable robustness; and every distilled number is a single training run, so we do not
rank methods whose intervals overlap.

\textbf{Open limits.} Rare-word and number/date robustness are not solved (capability
and measurement limits respectively). The online iterative direction is promising but
unproven at scale. Total compute is $\sim$45 GPU-hours.

\section{Conclusion}
\label{sec:conclusion}
Catastrophic failures in neural-codec TTS are not an intrinsic price of the paradigm:
best-of-$N$ ASR self-verification drives them to near-zero on three of four models (and
substantially reduces the fourth), and, across four models and three neural codecs, this
reduction is a broadly shared property, not a single-model quirk. A single distillation
pass makes the fix free at inference where it matters most, on the hard inputs that
dominate the failure budget. Offline preference optimization adds nothing
over supervised distillation; an online iterative variant is the natural next step for
the residual hard-input gap. The remaining honest gaps, one model whose result is
inconclusive under a generation-budget confound, a rare-word capability ceiling, and a
within-corpus difficulty control we did not run, are signposted for follow-up rather
than papered over.

\bibliographystyle{plainnat}
\bibliography{references}

\begin{thebibliography}{16}
\providecommand{\natexlab}[1]{#1}
\providecommand{\url}[1]{\texttt{#1}}
\expandafter\ifx\csname urlstyle\endcsname\relax
  \providecommand{\doi}[1]{doi: #1}\else
  \providecommand{\doi}{doi: \begingroup \urlstyle{rm}\Url}\fi

\bibitem[Atamanenko et~al.(2025)Atamanenko, Chalova, Coombes, Cope, Dang, Deng,
  Du, Ermolenko, Fan, Feng, Fichter, Filimonov, Fischer, Gibbs, Gusarova,
  Karpik, Kottner, Lee, Louie, Mai, Mamontov, Mao, Morshed, Poletaev, Radu,
  Semernia, Shingarev, Sivaraja, Skirko, Takhautdinov, Villahermosa, and
  Wang]{2507.21138v1}
Oleg Atamanenko, Anna Chalova, Joseph Coombes, Nikki Cope, Phillip Dang,
  Zhifeng Deng, Jimmy Du, Michael Ermolenko, Feifan Fan, Yufei Feng, Cheryl
  Fichter, Pavel Filimonov, Louis Fischer, Kylan Gibbs, Valeria Gusarova, Pavel
  Karpik, Andreas~Assad Kottner, Ian Lee, Oliver Louie, Jasmine Mai, Mikhail
  Mamontov, Suri Mao, Nurullah Morshed, Igor Poletaev, Florin Radu, Dmytro
  Semernia, Evgenii Shingarev, Vikram Sivaraja, Peter Skirko, Rinat
  Takhautdinov, Robert Villahermosa, and Jean Wang.
\newblock {TTS-1 Technical Report}.
\newblock arXiv:2507.21138 [cs.CL], 2025.
\newblock URL \url{https://arxiv.org/abs/2507.21138}.

\bibitem[Baba et~al.(2024)Baba, Nakata, Saito, and Saruwatari]{2409.09305v1}
Kaito Baba, Wataru Nakata, Yuki Saito, and Hiroshi Saruwatari.
\newblock {The T05 System for the VoiceMOS Challenge 2024: Transfer Learning
  from Deep Image Classifier to Naturalness MOS Prediction of High-Quality
  Synthetic Speech}.
\newblock arXiv:2409.09305 [cs.SD], 2024.
\newblock URL \url{https://arxiv.org/abs/2409.09305}.

\bibitem[Broukhim et~al.(2025)Broukhim, Shen, Ammanabrolu, and
  Weibel]{2511.13936v1}
Aaron Broukhim, Yiran Shen, Prithviraj Ammanabrolu, and Nadir Weibel.
\newblock {Preference-Based Learning in Audio Applications: A Systematic
  Analysis}.
\newblock arXiv:2511.13936 [cs.SD], 2025.
\newblock URL \url{https://arxiv.org/abs/2511.13936}.

\bibitem[{Canopy Labs}(2025)]{orpheus2025}
{Canopy Labs}.
\newblock {Orpheus: Towards Human-Sounding Speech (Orpheus-3B TTS)}.
\newblock Model release, HuggingFace \texttt{canopylabs/orpheus-3b-0.1-ft};
  code \url{https://github.com/canopyai/Orpheus-TTS}, 2025.
\newblock URL \url{https://huggingface.co/canopylabs/orpheus-3b-0.1-ft}.

\bibitem[Du et~al.(2024)Du, Wang, Chen, Shi, Lv, Zhao, Gao, Yang, Gao, Wang,
  Yu, Liu, Sheng, Gu, Deng, Wang, Zhang, Yan, and Zhou]{2412.10117v3}
Zhihao Du, Yuxuan Wang, Qian Chen, Xian Shi, Xiang Lv, Tianyu Zhao, Zhifu Gao,
  Yexin Yang, Changfeng Gao, Hui Wang, Fan Yu, Huadai Liu, Zhengyan Sheng, Yue
  Gu, Chong Deng, Wen Wang, Shiliang Zhang, Zhijie Yan, and Jingren Zhou.
\newblock {CosyVoice 2: Scalable Streaming Speech Synthesis with Large Language
  Models}.
\newblock arXiv:2412.10117 [cs.SD], 2024.
\newblock URL \url{https://arxiv.org/abs/2412.10117}.

\bibitem[Gorbatovski et~al.(2025)Gorbatovski, Shaposhnikov, Sinii, Malakhov,
  and Gavrilov]{2502.01237v3}
Alexey Gorbatovski, Boris Shaposhnikov, Viacheslav Sinii, Alexey Malakhov, and
  Daniil Gavrilov.
\newblock {The Differences Between Direct Alignment Algorithms are a Blur}.
\newblock arXiv:2502.01237 [cs.LG], 2025.
\newblock URL \url{https://arxiv.org/abs/2502.01237}.

\bibitem[Hussain et~al.(2025{\natexlab{a}})Hussain, Neekhara, Yang, Casanova,
  Ghosh, Desta, Fejgin, Valle, and Li]{2502.05236v2}
Shehzeen Hussain, Paarth Neekhara, Xuesong Yang, Edresson Casanova, Subhankar
  Ghosh, Mikyas~T. Desta, Roy Fejgin, Rafael Valle, and Jason Li.
\newblock {Koel-TTS: Enhancing LLM based Speech Generation with Preference
  Alignment and Classifier Free Guidance}.
\newblock arXiv:2502.05236 [cs.SD], 2025{\natexlab{a}}.
\newblock URL \url{https://arxiv.org/abs/2502.05236}.

\bibitem[Hussain et~al.(2025{\natexlab{b}})Hussain, Neekhara, Yang, Casanova,
  Ghosh, Fejgin, Langman, Desta, Tavabi, and Li]{2509.21718v1}
Shehzeen Hussain, Paarth Neekhara, Xuesong Yang, Edresson Casanova, Subhankar
  Ghosh, Roy Fejgin, Ryan Langman, Mikyas Desta, Leili Tavabi, and Jason Li.
\newblock {Align2Speak: Improving TTS for Low Resource Languages via ASR-Guided
  Online Preference Optimization}.
\newblock arXiv:2509.21718 [cs.AI], 2025{\natexlab{b}}.
\newblock URL \url{https://arxiv.org/abs/2509.21718}.

\bibitem[Jiang et~al.(2025)Jiang, Ren, Li, Ji, Zhang, Ye, Zhang, Jionghao,
  Yang, Zuo, Zhang, Liu, Yin, and Zhao]{2502.18924v4}
Ziyue Jiang, Yi~Ren, Ruiqi Li, Shengpeng Ji, Boyang Zhang, Zhenhui Ye, Chen
  Zhang, Bai Jionghao, Xiaoda Yang, Jialong Zuo, Yu~Zhang, Rui Liu, Xiang Yin,
  and Zhou Zhao.
\newblock {MegaTTS 3: Sparse Alignment Enhanced Latent Diffusion Transformer
  for Zero-Shot Speech Synthesis}.
\newblock arXiv:2502.18924 [eess.AS], 2025.
\newblock URL \url{https://arxiv.org/abs/2502.18924}.

\bibitem[Liu et~al.(2025)Liu, Fang, Hu, Zhang, Zhou, Zhang, Tu, Lin, Huang,
  Song, Li, and Tao]{2503.11701v1}
Shunyu Liu, Wenkai Fang, Zetian Hu, Junjie Zhang, Yang Zhou, Kongcheng Zhang,
  Rongcheng Tu, Ting-En Lin, Fei Huang, Mingli Song, Yongbin Li, and Dacheng
  Tao.
\newblock {A Survey of Direct Preference Optimization}.
\newblock arXiv:2503.11701 [cs.LG], 2025.
\newblock URL \url{https://arxiv.org/abs/2503.11701}.

\bibitem[Minixhofer et~al.(2024)Minixhofer, Klejch, and Bell]{2407.12707v3}
Christoph Minixhofer, Ond{\v{r}}ej Klejch, and Peter Bell.
\newblock {TTSDS {--} Text-to-Speech Distribution Score}.
\newblock arXiv:2407.12707 [eess.AS], 2024.
\newblock URL \url{https://arxiv.org/abs/2407.12707}.

\bibitem[{Sesame AI}(2025)]{csm2025}
{Sesame AI}.
\newblock {CSM: A Conversational Speech Generation Model}.
\newblock Model release, HuggingFace \texttt{sesame/csm-1b}; code
  \url{https://github.com/SesameAILabs/csm}, 2025.
\newblock URL \url{https://huggingface.co/sesame/csm-1b}.

\bibitem[Wang et~al.(2025)Wang, Zheng, Zhang, Zhang, Liao, and
  Wu]{2502.03128v1}
Yuancheng Wang, Jiachen Zheng, Junan Zhang, Xueyao Zhang, Huan Liao, and
  Zhizheng Wu.
\newblock {Metis: A Foundation Speech Generation Model with Masked Generative
  Pre-training}.
\newblock arXiv:2502.03128 [cs.SD], 2025.
\newblock URL \url{https://arxiv.org/abs/2502.03128}.

\bibitem[Xiao et~al.(2024)Xiao, Wang, Gan, Zhao, Li, Lei, He, Tuan, Chen,
  Jiang, Zhao, and Wu]{2410.15595v3}
Wenyi Xiao, Zechuan Wang, Leilei Gan, Shuai Zhao, Zongrui Li, Ruirui Lei,
  Wanggui He, Luu~Anh Tuan, Long Chen, Hao Jiang, Zhou Zhao, and Fei Wu.
\newblock {A Comprehensive Survey of Direct Preference Optimization: Datasets,
  Theories, Variants, and Applications}.
\newblock arXiv:2410.15595 [cs.AI], 2024.
\newblock URL \url{https://arxiv.org/abs/2410.15595}.

\bibitem[Ye et~al.(2025)Ye, Zhu, Chan, Wang, Tan, Lei, Peng, Liu, Jin, Dai,
  Lin, Chen, Du, Xue, Chen, Li, Xie, Kong, Guo, and Xue]{llasa2025}
Zhen Ye, Xinfa Zhu, Chi-Min Chan, Xinsheng Wang, Xu~Tan, Jiahe Lei, Yi~Peng,
  Haohe Liu, Yizhu Jin, Zheqi Dai, Hongzhan Lin, Jianyi Chen, Xingjian Du,
  Liumeng Xue, Yunlin Chen, Zhifei Li, Lei Xie, Qiuqiang Kong, Yike Guo, and
  Wei Xue.
\newblock {Llasa: Scaling Train-Time and Inference-Time Compute for Llama-based
  Speech Synthesis}.
\newblock arXiv:2502.04128 [eess.AS], 2025.
\newblock URL \url{https://arxiv.org/abs/2502.04128}.

\bibitem[Zhang et~al.(2025)Zhang, Zhang, Yang, Wang, Fan, and Wu]{2508.17229v2}
Junan Zhang, Xueyao Zhang, Jing Yang, Yuancheng Wang, Fan Fan, and Zhizheng Wu.
\newblock {Multi-Metric Preference Alignment for Generative Speech
  Restoration}.
\newblock arXiv:2508.17229 [cs.SD], 2025.
\newblock URL \url{https://arxiv.org/abs/2508.17229}.

\end{thebibliography}
\end{document}